\documentclass[11pt]{article}
\usepackage{a4wide}
\usepackage{authblk}
\usepackage{amsmath,amssymb,amsfonts,latexsym,stmaryrd}
\usepackage{mathtools}
\usepackage[mathscr]{eucal}
\usepackage{url}
\usepackage[PostScript=dvips]{diagrams}
\usepackage{stmaryrd}
\usepackage{amsthm}
\usepackage{tikz-cd}
\usepackage[framemethod=TikZ]{mdframed}

\newtheorem{theorem}{Theorem}

\theoremstyle{definition}

\mdfsetup{%
middlelinecolor=red,
middlelinewidth=2pt,
backgroundcolor=red!10,
roundcorner=10pt}

\newcommand{\ie}{\textit{i.e.}~}

\newcommand{\rarr}{\rightarrow}

\newcommand{\vphi}{\varphi}

\newcommand{\Real}{\mathbb{R}}
\newcommand{\Two}{\mathbf{2}}

\usepackage{colonequals}
\newcommand{\defeq}{\colonequals} 
\newcommand{\tuple}[1]{\mathopen{\langle}#1\mathclose{\rangle}} 

\newcommand{\fdec}    [3]{#1 \colon #2 \longrightarrow #3}

\newcommand{\setdef}  [2]{\left\{#1 \mid #2\right\}}  

\newcommand{\enset}   [1]{\mathopen{ \{ }#1\mathclose{ \} }} % {a,b,...z}

\newcommand{\RRpz}{\mathbb{R}_{\geq 0}}

\newcommand{\Mcomma}{\text{,}}

\newcommand{\Mdot}{\text{.}}
\newcommand{\Mand}{\quad\quad\text{ and }\quad\quad}

\newcommand{\M}{\mathcal{M}}
\newcommand{\XMO}{\tuple{X, \M, O}}

\newcommand{\Forall}[1]{\forall {#1}\boldsymbol{.}\;}

\newcommand{\dg}{\delta^g}
\newcommand{\ve}{\vect{v}^{e}}

\newcommand{\IM}{\vect{M}}
\newcommand{\vect}[1]{\mathbf{#1}}
\newcommand{\rr}{\mathbf{r}}
\newcommand{\vv}{\mathbf{v}}
\newcommand{\av}{\vect{a}}
\newcommand{\vone} {\vect{1}}
\newcommand{\vzero}{\vect{0}}
\newcommand{\bb}{\vect{b}}
\newcommand{\dd}{\vect{d}}
\newcommand{\RR}  {\mathbb{R}}
\newcommand{\weight}[1]{w(#1)}

\def\stress{\textit}

\newcommand{\CF} {\mathsf{CF}}
\newcommand{\NCF}{\mathsf{NCF}}

\newcommand{\textLP}{\texttt}

\newcommand{\rowsp}{\hspace{1em}}

\begin{document}

\title{Classical logic, classical probability, and quantum mechanics}
\author{Samson Abramsky\thanks{samson.abramsky@cs.ox.ac.uk}~}
\affil{Department of Computer Science, University of Oxford}
\date{}
\maketitle

\begin{abstract}
We give an overview and conceptual discussion of some of our results on contextuality and non-locality. We focus in particular on connections with the work of Itamar Pitowsky on correlation polytopes, Bell inequalities, and Boole's ``conditions of possible experience''.
\end{abstract}

\section{Introduction}
One of Itamar Pitowsky's most celebrated contributions to the foundations of quantum mechanics was his work on \emph{correlation polytopes} \cite{pitowsky1989quantum,pitowsky1991correlation,pitowsky1994george}, as a general perspective on Bell inequalities. He related these to Boole's ``conditions of possible experience'' \cite{boole1862theory}, and emphasized the links between correlation polytopes and classical logic.

My own work on the sheaf-theoretic approach to non-locality and contextuality with Adam Brandenburger \cite{AbramskyBrandenburger}, on logical Bell inequalities with Lucien Hardy \cite{AbramskyHardy:LogicalBellIneqs}, on robust constraint satisfaction with Georg Gottlob and Phokion Kolaitis \cite{abramsky2013robust}, and on the contextual fraction with Rui Soares Barbosa and Shane Mansfield \cite{abramsky2017contextual}, is very much in a kindred spirit with Pitowsky's pioneering contributions. I will survey some of this work, making a number of  comparisons and contrasts with Pitowsky's ideas.

\section{Boole's ``conditions of possible experience''}
We quote Pitowsky's pellucid summary \cite[p.~100]{pitowsky1994george}:
\begin{quotation}
\textit{Boole's problem is simple: we are given rational
numbers which indicate the relative frequencies of certain events. If no logical
relations obtain among the events, then the only constraints imposed on these
numbers are that they each be non-negative and less than one. If however, the
events are logically interconnected, there are further equalities or inequalities
that obtain among the numbers. The problem thus is to determine the
numerical relations among frequencies, in terms of equalities and inequalities,
which are induced by a set of logical relations among the events. The equalities
and inequalities are called ``conditions of possible experience''.}
\end{quotation}

More formally, we are given some basic events $E_1, \ldots , E_n$, and some boolean functions $\vphi_1, \ldots , \vphi_m$ of these events.
Such a function $\vphi$ can be described  by a propositional formula in the variables $E_1, \ldots , E_n$. 

Suppose further that we are given probabilities $p(E_i)$, $p(\vphi_j)$ of these events.
\begin{mdframed}[align=center,userdefinedwidth=31em]
Question: What numerical relationships between the probabilities \\
can we infer 
from the logical relationships between the events?
\end{mdframed}

Pitowksy's approach is to define a \emph{correlation polytope} $c(\vec{E}, \vec{\vphi})$ induced by the given events, and to characterize the ``conditions of possible experience'' as the facet-defining inequalities for this polytope. He emphasizes the computational difficulty of obtaining these inequalities, and gives no direct characterization other than a brute-force computational approach. Indeed, one of his main results is the NP-completeness of an associated problem, of determining membership of the polytope.
We shall return to these ideas later.

The point we wish to make now is that it is possible to give a very direct answer to Boole's question, which explicitly relates logical consistency conditions to arithmetical relationships.
This insight is the key observation in \cite{AbramskyHardy:LogicalBellIneqs}.

With notation as above, suppose that the formulas $\vphi_j$ are not simultaneously satisfiable. This means that $\bigwedge_{i=1}^{m-1} \vphi_i \rarr \neg \vphi_m$, or by contraposition and De Morgan's law:
\[ \vphi_m \; \rarr \; \bigvee_{i=1}^{m-1} \neg \vphi_i . \]
Passing to probabilities, we infer:
\[ p(\vphi_m) \; \leq \; p(\bigvee_{i=1}^{m-1} \neg \vphi_i) \; \leq \; \sum_{i=1}^{m-1} p(\neg \vphi_i) \; = \; \sum_{i=1}^{m-1} (1 - p(\vphi_i)) \; = (m-1) -  \sum_{i=1}^{m-1} p(\vphi_i) . \]
The first inequality is the monotonicity of probability, \ie $E \subseteq E'$ implies $p(E) \leq p(E')$, while the second is the subadditivity of probability measures.\footnote{Also known as Boole's inequality \cite{Bonf}.}
Collecting terms, we obtain:
\begin{equation}
\label{LBell} 
\sum_{i=1}^{m} p(\vphi_i) \; \leq \; m-1 . 
\end{equation}
Thus a consistency condition on the formulas $\vphi_i$ implies a non-trivial arithmetical inequality on the probabilities $p(\vphi_i)$.

\subsection{An example: the Bell table}
\label{bellsec}

We can directly apply the above inequality to deduce a version of Bell's theorem \cite{bell1964einstein}. Consider the following table.

\begin{center}
\begin{tabular}{cc|ccccc}
A & B & $(0, 0)$ & $(1, 0)$ & $(0, 1)$ & $(1, 1)$  &  \\ \hline
$a$ & $b$ & $1/2$ & $0$ & $0$ & $1/2$ & \\
$a$ & $b'$ & $3/8$ & $1/8$ & $1/8$ & $3/8$ & \\
$a'$ & $b$ & $3/8$ & $1/8$ & $1/8$ & $3/8$ &  \\
$a'$ & $b'$ & $1/8$ & $3/8$ & $3/8$ & $1/8$ &
\end{tabular}
\end{center}

Here we have two agents, Alice and Bob. Alice can choose from the measurement settings $a$ or $a'$, and Bob can choose from $b$ or $b'$. These choices correspond to the rows of the table. The columns correspond to the joint outcomes for a given choice of settings by Alice and Bob, the two possible outcomes for each individual measurement being represented by $0$ and~$1$.
The numbers along each row specify a probability distribution on these joint outcomes. Thus for example, the entry in row 2, column 2 of the table says that when Alice chooses setting $a$ and Bob chooses setting $b'$, then with probability $1/8$, Alice obtains a value of $1$, and Bob obtains a value of $0$.

A standard version of Bell's theorem uses the probability table given above. This table can be realized in quantum mechanics, e.g.~by a Bell state
%, written in the $Z$ basis as
\[ \frac{| 00  \rangle \; + \; | 11 \rangle}{\sqrt{2}} , \]
subjected to spin measurements in the   $XY$-plane of the Bloch sphere, at a relative angle of $\pi/3$. See the supplemental material in \cite{abramsky2017contextual} for details.

\subsubsection*{Logical analysis of the Bell table}
We now pick out a subset of the elements of each row of the table, as indicated in the following table.
\begin{center}
\begin{tabular}{cc|ccccc}
A & B & $(0, 0)$ & $(1, 0)$ & $(0, 1)$ & $(1, 1)$  &  \\ \hline
$a$ & $b$ & { \fcolorbox{gray}{yellow}{1/2}} & $0$ & $0$ & { \fcolorbox{gray}{yellow}{1/2}} & \\
$a$ & $b'$ &  { \fcolorbox{gray}{yellow}{3/8}} & $1/8$ & $1/8$ & { \fcolorbox{gray}{yellow}{3/8}} & \\
$a'$ & $b$ & { \fcolorbox{gray}{yellow}{3/8}} & $1/8$ & $1/8$ & { \fcolorbox{gray}{yellow}{3/8}} &  \\
$a'$ & $b'$ & $1/8$ & { \fcolorbox{gray}{yellow}{3/8}} & { \fcolorbox{gray}{yellow}{3/8}} & $1/8$ &
\end{tabular}
\end{center}

We can think of basic events $E_a$, $E_{a'}$, $E_{b}$, $E_{b'}$, where e.g. $E_a$ is the event that the quantity measured by $a$ has the value $0$.
Note that, by our assumption that each measurement has two possible outcomes\footnote{And, implicitly, the assumption that each quantity has a definite value, whether we measure it or not.}, $\neg E_a$ is the event that $a$ has the value $1$. Writing simply $a$ for $E_a$ etc., 
%If we read $0$ as true and $1$ as false, 
the highlighted positions in the table are represented by the following propositions:
\[ \begin{array}{rcccccccc}
\vphi_1 & = &  a \wedge b & \vee & \neg a \wedge \neg b & = & a & \leftrightarrow & b \\
\vphi_2 & = & a \wedge b' & \vee & \neg a \wedge \neg b' & = &  a & \leftrightarrow & b' \\
\vphi_3 & = & a' \wedge b & \vee & \neg a' \wedge \neg b & = &  a' & \leftrightarrow & b \\
\vphi_4 & = & \neg a' \wedge b' & \vee & a' \wedge \neg b' & = & a' & \oplus & b' .
\end{array}
\]
The first three propositions pick out  the correlated outcomes for the variables they refer to; the fourth the anticorrelated outcomes.
These propositions are easily seen to be contradictory. Indeed, starting with
$\vphi_4$, we can replace $a'$ with $b$ using $\vphi_3$, $b$ with $a$ using $\vphi_1$, and $a$ with $b'$ using $\vphi_2$, to obtain $b' \oplus b'$, which is obviously unsatisfiable.

We see from the table that $p(\vphi_1) = 1$, $p(\vphi_i) = 6/8$ for $i = 2, 3, 4$.
Hence the violation of the Bell inequality~(\ref{LBell}) is $1/4$.

We may note that the logical pattern shown by this jointly contradictory family of propositions underlies the familiar CHSH correlation function.

\subsection{The general form}

We can generalize the inequality~(\ref{LBell}). Given a family of propositions $\Phi \, = \, \{ \vphi_i \}$, we say it is \emph{$K$-consistent} if the size of the largest consistent subfamily of $\Phi$ is $K$.\footnote{The size of a family $\{ \vphi_j \}_{j \in J}$ is the cardinality of $J$.
Note that repetitions are allowed in the family.} 
If a family $ \{ \vphi_i \}_{i=1}^{m}$ is not simultaneously satisfiable, then it must be $K$-consistent for some $K < m$.

\begin{theorem}
Suppose that we have a $K$-consistent family $\{ \vphi_i \}_{i=1}^{m}$ over the basic events $E_1, \ldots , E_n$. For any probability distribution on the set $\mathbf{2}^{\vec{E}}$ of truth-value assignments to the $E_i$, with induced probabilities $p(\vphi_i)$ for the events $\vphi_i$, we have:
\begin{equation}
\label{LBell2}
\sum_{i=1}^{m} p(\vphi_i) \; \leq \; K .
\end{equation}
\end{theorem}

See \cite{AbramskyHardy:LogicalBellIneqs} for the (straightforward) proof.
Note that the basic inequality~(\ref{LBell}) is a simple consequence of this result.

We thus have a large class of inequalities arising directly from logical consistency conditions. As shown in \cite{AbramskyHardy:LogicalBellIneqs}, even the basic form~(\ref{LBell}) is sufficient to derive the main no-go results in the literature, including the Hardy, GHZ and Kochen-Specker ``paradoxes''.
 
More remarkably, as we shall now go on to see, this set of inequalities is \emph{complete}. That is, \emph{every} facet-defining inequality for the local, or more generally non-contextual polytopes, is equivalent to one of the form~(\ref{LBell2}). In this sense, we have given a complete answer to Boole's question.

The following quotation from Pitowsky suggests that he may have envisaged the possibility of such a result \cite[p.~413]{pitowsky1991correlation}:
\begin{quotation}
\textit{In fact, all facet inequalities for $c(n)$ should follow from ``Venn diagrams'', that is, the possible relations among $n$ events in a probability space.}
\end{quotation}

\section{From correlation polytopes to non-contextual polytopes}

We continue with the notation $E_1, \ldots , E_n$ for basic events, and  $\vphi_1, \ldots , \vphi_m$ for boolean combinations of these events.
We write $\Two  :=  \{ 0, 1 \}$.

Pitowsky defines the correlation polytope $c(\vec{E}, \vec{\vphi}) \subseteq \Real^{n+m}$ as follows. Each assignment $s \in \Two^{\vec{E}}$ of $0$ or $1$ to the basic events $E_i$ extends to a truth assignment for the formulae $\vphi_j$ by the usual truth-table method, and hence determines a $0/1$-vector $v_s$ in $\Real^{n+m}$. The polytope $c(\vec{E}, \vec{\vphi})$ is defined to be the convex hull of the set of vectors $v_s$, for $s \in \Two^{\vec{E}}$.

In particular, Pitowsky focusses almost exclusively on the case where the formulae are pairwise conjunctions $\vphi_{ij} \, := \, E_i \wedge E_j$. One effect of restricting to this case is that $m$ is bounded quadratically by $n$. We shall return to this point when we discuss complexity issues.

The main computational problem that Pitowsky focusses on is the following:
\begin{mdframed}[align=center,userdefinedwidth=18em]
Instance: a rational vector $v \in \mathbf{Q}^{n+m}$. \\
Question: is $v$ in $c(\vec{E}, \vec{\vphi})$?
\end{mdframed}
For the case where $\vec{\vphi}$ is the set of all pairwise conjunctions $\vphi_{ij}$ of basic events, Pitowsky shows that this problem is NP-complete.

We can see that the (quite standard) version of Bell's theorem described in section~\ref{bellsec} amounts to the statement that the vector 
\[ v = [p_a \rowsp p_{a'} \rowsp p_{b} \rowsp p_{b'} \rowsp 1 \rowsp 6/8 \rowsp 6/8 \rowsp 6/8]^T \]
is not in the correlation polytope $c(E_a, E_{a'}, E_{b}, E_{b'}, \vphi_1, \vphi_2, \vphi_3, \vphi_4)$. Here $p_a$ is the probability of the event $E_a$. This can be calculated as a marginal from the first or second rows of the Bell table, as $p_a = 1/2$. The fact that we get the same answer, whether we use the first or second row, is guaranteed by the \emph{no-signalling condition} \cite{jordan1983quantum,ghirardi1980general}, which will hold for all such tables which can be generated using quantum states and observables \cite{AbramskyBrandenburger}. Similarly for the other basic events.

We make a number of comments:
\begin{itemize}
\item Firstly, notice that the Bell table as given in section~\ref{bellsec} corresponds to the data arising directly from a Bell experiment. This and similar tables should be seen as setting the standard for what we are trying to describe.
\item From this point of view, we notice firstly that to get a natural correspondence between these tables and correlation polytopes, we need to consider propositions beyond pairwise conjunctions. This makes a huge difference as regards the size of problem instances. While there are only $O(n^2)$ pairwise conjunctions on $n$ variables, there are $2^{2^n}$ distinct boolean functions.
\item At the same time, there is redundancy in the correlation polytope representation, since thanks to the no-signalling condition, the probabilities of the basic events can be calculated as marginals from the probabilities of the joint outcomes.
\item The most important point is that there is a \emph{structure} inherent in the Bell table and its generalizations, which is ``flattened out'' by the correlation polytope representation.
Capturing this structure explicitly can lead to deeper insights into the non-classical phenomena arising in quantum mechanics.
\end{itemize}

Taking up the last point, we quote again from Pitowsky's admirably clear exposition \cite[p.~111--112]{pitowsky1994george}:
\begin{quotation}
\textit{One of the major
purposes of quantum mechanics is to organize and predict the relative frequencies of events
observable in various experiments --- in particular the cases where Boole's
conditions are violated. For that purpose a mathematical formalism has been
invented which is essentially a new kind of probability theory. It uses no
concept of `population' but rather a primitive concept of `event' or more
generally `observable' (which is the equivalent of the classical `random
variable'). 
In addition, to every particular physical system (which can be one
`thing' --- an electron, for example --- or consists of a few `things') the theory
assigns a state. The state determines the probabilities for the events or, more
generally, the expectations of the observables. What this means operationally
is that if we have a source of physical systems, all in the same state, then the
relative frequency of a given event will approach the value of the probability,
which is theoretically determined by the state.}

\textit{For certain families of events the theory stipulates that they are commeasurable. 
This means that, in every state, the relative frequencies of all these events
can be measured on one single sample. For such families of events, the rules
of classical probability --- Boole's conditions in particular --- are valid. Other
families of events are not commeasurable, so their frequencies must be
measured in more than one sample. The events in such families nevertheless
exhibit logical relations (given, usually, in terms of algebraic relations among
observables). But for some states, the probabilities assigned to the events
violate one or more of Boole's conditions associated with those logical
relations.}
\end{quotation}

The point we would like to emphasize is that tables such as the Bell table in section~\ref{bellsec} can --- and do --- arise from experimental data, without presupposing any particular physical  theory.
This data \emph{does} clearly involve concepts from classical probability. In particular, for each set $C$ of ``commeasurable events'', there is  a sample space, namely the set of possible joint outcomes of measuring all the observables in $C$. Moreover, there is a well-defined probability distribution on this sample space.

Taking the Bell table for illustration, the ``commeasurable sets'', or \emph{contexts}, are the sets of measurements labelling the rows of the table. The sample space for a given row, with measurement $\alpha$ by Alice and $\beta$ by Bob, is the set of joint outcomes $(\alpha=x,\beta=y)$ with $x, y \in \{ 0, 1 \}$. The probabilities assigned to these joint outcomes in the table form a (perfectly classical) probability distribution on this sample space.

Thus the structure of the table as a whole is a family of probability distributions, each defined on a different sample space. However, these sample spaces are not completely unrelated. They overlap in sharing common observables, and they ``agree on overlaps'' in the sense that they have consistent marginals. This is exactly the force of the no-signalling condition.

The nature of the ``non-classicality'' of the data tables generated by quantum mechanics (and also elsewhere, see e.g.~\cite{abramsky2013relational}), is that there is no distribution over the global sample space of outcomes  for \emph{all} observables, which recovers the empirically accessible data in the table by marginalization. This is completely equivalent to a  statement of the more traditional form: ``there is no local or non-contextual hidden variable model'' (or, in currently fashionable terminology: there is no ``ontological model'' of a certain form).

Our preferred slogan for this state of affairs is:
\begin{mdframed}[align=center,userdefinedwidth=35em]
Contextuality (with non-locality as a special case) arises when we have a family of data which is locally consistent, but globally inconsistent.
\end{mdframed}

\subsection{Formalization}

We briefly summarise the framework introduced in \cite{AbramskyBrandenburger}, and extensively developed subsequently.
The main objects of study are \emph{empirical models}: tables of data, specifying probability distributions over the joint outcomes of specified sets of compatible measurements.
These can be thought of as statistical data obtained from some experiment or as the observations predicted by some theory.

A \emph{measurement scenario} is an abstract description of a particular experimental setup.
It consists of a triple $\XMO$ where:
$X$ is a finite set of measurements;
$O$ is a finite set of outcome values for each measurement;
and $\M$ is a set of subsets of $X$.
Each $C \in \M$ is called a  \emph{measurement context}, and represents a set of measurements that can be performed together.

%\begin{figure}
%\begin{center}
%\ABscheme %\includegraphics[scale=0.8]{fig_1}
%\end{center}
%\caption{\label{fig:Bell-scenario-scheme} Schematic representation of the experimental setup for the $(2,2,2)$ Bell scenario. %{\color{red}A luxury figure\dots}
%}
%\end{figure}

Examples of measurement scenarios include multipartite Bell-type scenarios familiar from discussions of nonlocality, Kochen--Specker configurations, measurement scenarios associated with qudit stabiliser quantum mechanics, and more.
For example, the Bell scenario from section~\ref{bellsec},
%depicted in Figure~\ref{fig:Bell-scenario-scheme},
where two experimenters, Alice and Bob,
can each choose between performing one of two different measurements, $a$ or $a'$ for Alice and $b$ or $b'$ for Bob,
obtaining one of two possible outcomes, is represented as follows:
\begin{align*}
X &= \enset{a,a',b,b} \qquad\qquad O = \enset{0,1} \\
\M &= \enset{ \enset{a,b} , \enset{a,b'} , \enset{a',b}, \enset{a', b'} }
\Mdot
\end{align*}

Given this description of the experimental setup, then either performing repeated runs of such experiments with varying choices of measurement context and recording the frequencies of the various outcome events, or calculating theoretical predictions for the probabilities of these outcomes, results in a
probability table like that in section~\ref{bellsec}.

Such data is formalised as an \emph{empirical model} for the given measurement scenario $\XMO$.
For each valid choice of measurement context,
it specifies the probabilities of obtaining the corresponding joint outcomes.
That is, it is a family $\{e_C\}_{C \in \M}$
where each  $e_C$ is a probability distribution on the set $O^C$ of functions assigning an outcome in $O$ to each measurement in $C$ (the rows of the probability table).

We require that the marginals of these distributions agree whenever contexts overlap, \ie
\[\Forall{C, C' \in \M} e_C|_{C \cap C'} = e_{C'}|_{C \cap C'} \Mcomma\]
where the notation $e_C|_{U}$ with $U \subseteq C$ stands for marginalisation of probability distributions (to `forget' the outcomes of some measurements): for $t \in O^U$, $e_C|_{U}(t) \defeq \sum_{s \in O^C, s|_U = t}e_C(t)$.
The requirement of \emph{compatibility of marginals} is a generalisation of the usual \emph{no-signalling} condition, and is satisfied in particular by all empirical models arising from quantum predictions \cite{AbramskyBrandenburger}.

%{\color{orange}MIXING HERE INSTEAD OF BELOW?}

An empirical model is said to be \emph{non-contextual} if this family of distributions
can be obtained as the marginals of a single probability distribution on global assignments of outcomes to all measurements,
\ie a distribution $d$ on $O^X$ (where $O^X$ acts as a canonical set of deterministic hidden variables) such that $\Forall{C \in \M} d|_C = e_C$.
Otherwise, it is said to be \emph{contextual}.
%{\color{orange}
Equivalently \cite{AbramskyBrandenburger}, contextual empirical models are those which have no realisation by factorisable hidden variable models; thus for Bell-type measurement scenarios contextuality specialises to the usual notion of \emph{nonlocality}.

Noncontextuality characterizes classical behaviours.
One way to understand this is that it reflects a situation in which the physical system being measured exists at all times in a definite state assigning outcome values to all properties that can be measured.
Probabilistic behaviour may still arise, but only via stochastic mixtures or distributions on  these global assignments.
This may reflect an averaged or aggregate behaviour,
or an epistemic limitation on our knowledge of the underlying global assignment.

\subsection{The non-contextual polytope}

Suppose we are given a measurement scenario  $\XMO$. 
Each global assignment $g \in O^X$ induces a deterministic empirical model $\dg$:
\[ \dg_C(s) = \left\{ \begin{array}{ll}
1, & g |_C = s \\
0 & \mbox{otherwise.}
\end{array}
\right.
\]
Note that $g |_C$ is the function $g$ restricted to $C$, which is a subset of its domain $X$.

We have the following result from \cite[Theorem 8.1]{AbramskyBrandenburger}:
\begin{theorem}
\label{ncdetth}
 An empirical model $\{ e_C \}$ is non-contextual if and only if it can be written as a convex combination $\sum_{j \in J} \mu_j \delta^{g_j}$ where $g_j \in O^X$ for each $j \in J$.
This means that for each $C \in \M$, 
\[ e_C = \sum_j \mu_j \delta^{g_j}_{C} . \]
\end{theorem}

Given a scenario $\XMO$, define $m \defeq \sum_{C \in \M} |O^C| = |\setdef{\tuple{C,s}}{C\in\M, s \in O^C}|$ to be the number of joint outcomes as we range over contexts.
For example, in the case of the Bell table there are four contexts, each with four possible outcomes, so $m=16$.
We can regard an empirical model $\{ e_C \}$ over $\XMO$ as a real vector $\ve \in \Real^m$, with  $\ve[\tuple{C,s}] = e_C(s)$.

We define the \emph{non-contextual polytope} for the scenario $\XMO$ to be the convex hull of the set of  deterministic models $\dg$, where $g$ ranges over $O^X$.
By the preceding theorem, this is exactly the set of non-contextual models.

Thus we have captured the question as to whether an empirical model is contextual in terms of membership of a polytope. The facet inequalities for this family of polytopes, as we range over measurement scenarios, give a general notion of Bell inequalities.

As explained in \cite{AbramskyHardy:LogicalBellIneqs}, a complete finite set of rational facet inequalities for the contextual polytope over a measurement scenario can be computed using Fourier-Motzkin elimination. This procedure is doubly-exponential in the worst case, but standard optimizations reduce this to a single exponential. Despite this high complexity, Fourier-Motzkin elimination is widely used in computer-assisted verification and polyhedral computation \cite{strichman2002solving,christof1997porta}.

\subsection{Completeness of logical Bell inequalities}

Suppose we are given a scenario $\XMO$. A rational inequality  is given by a rational vector $\rr$ and a rational number $r$. An empirical model $\vv$ satisfies this inequality if $\rr \cdot \vv \leq r$.
Two inequalities are equivalent if they are satisfied by the same empirical models.

\begin{theorem}
\label{ratineqequivlbellth}
A rational inequality is satisfied by all non-contextual empirical models over $\XMO$ if and only if it is equivalent to a logical Bell inequality of the form~(\ref{LBell2}).
\end{theorem}

For the proof, see \cite{AbramskyHardy:LogicalBellIneqs}. Combining this result with the previous observations, we obtain:

\begin{theorem}
\label{compth}
The polytope of non-contextual empirical models over any scenario $\XMO$ is determined by a finite set of logical Bell inequalities. Moreover, these inequalities can be obtained effectively from the scenario.
Thus an empirical model over any scenario is contextual if and only if it violates one of finitely many logical Bell inequalities.
\end{theorem}

It is worth reflecting on the conceptual import of these results. They are saying that the scope for non-classical correlations available to quantum mechanics, or any other physical theory, arises entirely from families of events which are locally consistent, but globally inconsistent. Thus non-classical probabilistic behaviour rests on this kind of logical structure.
%See \cite{kishida2016logic} for a study of a ``logic of local inference'' suggested by these ideas.

\section{The contextual fraction}

We now turn to computational aspects. 
Fix a measurement scenario $\XMO$.
Let $n \defeq |O^X|$ be the number of global assignments $g$,
and $m \defeq |\setdef{\tuple{C,s}}{C\in\M, s \in O^C}|$ be the number of local assignments ranging over contexts.
The \emph{incidence matrix} \cite{AbramskyBrandenburger} $\IM$ is an $m \times n$ $(0,1)$-matrix that records the restriction relation between global and local assignments:
\[
\IM[\tuple{C,s},g] \defeq
\begin{cases}
 1 \;\text{ if $g|_C = s$;}
\\ 
 0 \;\text{ otherwise.}
\end{cases}
\]
As already explained, an empirical model $e$ can be represented as a vector $\ve \in \RR^m$, with the component $\ve[\tuple{C,s}]$ recording the probability given by the model to the assignment $s$ at the measurement context $C$, $e_C(s)$. 
This vector is a flattened version of the table used to represent the empirical model.
The columns of the incidence matrix, $\mathbf{M}[-,g]$, are the vectors corresponding to the (non-contextual) deterministic models $\delta^{g}$ obtained from global assignments $g \in O^X$.
Recall that every non-contextual model can be written as a convex combination of these.
A probability distribution on global assignments can be represented as a vector $\dd \in \RR^n$ with non-negative components, and then the corresponding non-contextual model is represented by the vector $\IM \, \dd$.
So a model $e$ is non-contextual if and only if there exists $\dd \in \RR^n$ such that:
\[\IM \, \dd \,=\, \ve \Mand \dd \geq \mathbf{0} \Mdot\]

It is also natural to consider a relaxed version of this question, which leads us to the contextual fraction.

Given two empirical models $e$ and $e'$ on the same measurement scenario
and $\lambda \in [0,1]$, we define the empirical model $\lambda e + (1-\lambda) e'$ by taking the convex sum of probability distributions at each context.
Compatibility is preserved by this convex sum, hence it yields a well-defined empirical model.

A natural question to ask is: what fraction of a given empirical model $e$ admits a non-contextual explanation?
This approach enables a refinement of the binary notion of contextuality \textit{vs} non-contextuality into a quantitative grading.
%That is,
Instead of asking for a probability distribution on global assignments
that marginalises to the empirical distributions at each context,
we ask only for a subprobability distribution\footnote{A subprobability distribution on a set $S$ is a map $\fdec{b}{S}{\RRpz}$ with finite support and $\weight{b} \leq 1$, where $\weight{b} \defeq \sum_{s \in S}b(s)$ is called its weight. The set of subprobability distributions on $S$ is ordered pointwise: $b'$ is a subdistribution of $b$ (written $b' \leq b$) whenever $\Forall{s \in S}b'(s) \leq b(s)$.} $b$ on global assignments $O^X$
that marginalises at each context to a subdistribution of the empirical data, thus explaining a fraction of the events,
\ie $\Forall{C \in \M} b|_C \leq e_C$.
Equivalently, we ask for a convex decomposition
\begin{equation}\label{eq:convexdec}
  e = \lambda e^{NC} + (1-\lambda) e'
\end{equation}
where $e^{NC}$ is a non-contextual model and $e'$ is another (no-signalling) empirical model.
The maximum 
weight of such a global subprobability distribution, or the maximum value of $\lambda$ in such a decomposition\footnote{Note that such a maximum exists, \ie that the supremum is attained. This follows from the Heine--Borel and extreme value theorems since the set of such $\lambda$ is bounded and closed.},
is called the \emph{non-contextual fraction} of $e$,
by analogy with the  \emph{local fraction} previously introduced for models on Bell-type scenarios
\cite{ElitzurPopescuRohrlich1992:QuantumNonlocalityForEachPairInAnEnsemble}.
We denote it by $\NCF(e)$, and the contextual fraction by $\CF(e) \defeq 1 - \NCF(e)$.

The notion of contextual fraction in general scenarios was introduced in \cite{AbramskyBrandenburger},
where it was proved that a model is non-contextual if and only if its non-contextual fraction is $1$.

A global subprobability distribution is represented by a vector $\bb \in \RR^n$ with non-negative components, its weight being given by the dot product $\vone \cdot \bb$,
where $\vone \in \RR^n$ is the vector whose $n$ components are each $1$. The following LP thus calculates the non-contextual fraction of an empirical model $e$:
\begin{equation}\label{LP:quantifying}
\begin{alignedat}{3}
&\textLP{Find }       \;\;&&  \bb \in \RR^n
\\
&\textLP{maximising }   \;\;&& \vone \cdot \bb
\\
&\textLP{subject to } \;\;&& \IM \, \bb \,\leq\, \ve
\\
&\textLP{and }        \;\;&& \bb \,\geq\, \vzero
& \Mdot
\end{alignedat}
\end{equation}

An \emph{inequality} for a scenario $\tuple{X,\M,O}$ is given by
a vector $\av \in \RR^m$ of real coefficients indexed by local assignments
$\tuple{C,s}$, %(\ie $\tuple{C,s}$ with $C \in \M$ and $s \in O^C$),
and a bound $R$.
For a model $e$, the inequality reads
$\av \cdot \ve  \, \leq \, R$,
where
\[\av \cdot \ve \; = \; \sum_{C \in \M, s \in O^C} \, \av[\tuple{C,s}] \, e_C(s) \Mdot \]
Without loss of generality, we can take $R$ to be non-negative (in fact, even $R = 0$)
as any inequality is equivalent to one of this form.
We call it a \emph{Bell inequality} if it is satisfied by every non-contextual model.
If, moreover, it is saturated by some non-contextual model, the Bell inequality is said to be \emph{tight}.
A Bell inequality establishes a bound for the value of $\av \cdot \ve$ amongst non-contextual models $e$.
For more general models, this quantity is limited only
by the algebraic bound\footnote{We will consider only inequalities satisfying $R < \|\av\|$, which excludes inequalities trivially satisfied by all models, and avoids cluttering the presentation with special caveats about division by 0.}
\[\|\av\| \defeq \sum_{C \in \M}\max\setdef{\av[\tuple{C,s}]}{s\in O^C} \Mdot\]
The \emph{violation} of a Bell inequality $\tuple{\av,R}$ by a model $e$ is 
$\max\enset{0,\av \cdot \ve - R}$.
However, it is useful to normalise this value by the maximum possible violation
in order to give a better idea of the \stress{extent} to which the model violates the inequality.
The \emph{normalised violation}
of the Bell inequality by the model $e$ is 
\[\frac{\max\enset{0,\av \cdot \ve - R}}{\|\av\| - R} \Mdot\]

\begin{theorem}
Let $e$ be an empirical model.
\begin{enumerate}
%[label=(\roman)]
\item\label{item:atmost} The normalised violation by $e$ of any Bell inequality is at most $\CF(e)$;
\item\label{item:ineq} if $\CF(e) > 0$, this bound is attained, \ie there exists a Bell inequality whose normalised violation by $e$ is $\CF(e)$;
\item\label{item:tight} moreover, for any decomposition of the form
$e = \NCF(e) e^{NC} + \CF(e)e^{SC}$,
this Bell inequality is tight at the non-contextual model $e^{NC}$ (provided $\NCF(e)>0$)
and maximally violated at the strongly contextual model $e^{SC}$.
\end{enumerate}
\end{theorem}

The proof of this result is based on the Strong Duality theorem of linear programming \cite{DantzigThapa:LinearProgramming2}.
This provides an LP method of calculating a witnessing Bell inequality for any empirical model $e$.
For details, see the supplemental material in \cite{abramsky2017contextual}.

\section{Remarks on complexity}

We now return to the issue of the complexity of deciding whether a given empirical model is contextual.

It will be useful to consider the class of $(n,k,2)$ Bell scenarios. In these scenarios there are $n$ agents, each of whom has a choice of $k$ measurement settings, and all measurements have 2 possible outcomes. This gives rise to a measurement scenario $\XMO$ where $|X| = nk$. Each $C \in \M$ consists of $n$ measurements, one chosen by each agent. Thus $|\M| = k^n$. For each context $C$, there are $2^n$ possible outcomes.

An empirical model for an $(n,k,2)$ Bell scenario is thus given by a vector of $k^n 2^n$ probabilities. Thus the size of instances is exponential in the natural parameter $n$.
This is the real obstacle to tractable computation as we increase the number of agents.

Given an empirical model $\ve$ as an instance, we can use the linear program~(\ref{LP:quantifying}) given in the previous section to determine if it is contextual. The size of the linear program is determined by the incidence matrix, which has dimensions $p \times q$, where $p$ is the dimension of $\ve$, and $q = 2^{nk}$.

If we treat $n$ as the complexity parameter, and keep $k$ fixed, then $q = O(s^k)$, where $s$ is the size of the instance. Thus the linear program has size polynomial in the size of the instance, and membership of the non-contextual polytope can be decided in time polynomial in the size of the instance. 
%The same analysis applies if we treat $k$ as the complexity parameter, and keep $n$ fixed.

There is  an interesting contrast with Pitowsky's results on the NP-completeness of deciding membership in the correlation polytope of all binary conjunctions of basic events.
In Pitowsky's case, the size of instances for these special forms of correlation polytope is polynomial in the natural parameter, which is the number of basic events. If we consider the correlation polytopes which would  correspond directly to empirical models, the same argument as given above would apply: the instances would have exponential size in the natural parameter, while membership in the polytope could be decided by linear programming in time polynomial in the instance size.

We can also consider the situation where we fix $n$, and treat $k$ as the complexity parameter. In this case, note that the size of instances are polynomial in $k$, while the size of the incidence matrix is exponential in $k$. Thus linear programming does not help.
In fact, this seems to be the case that Pitowsky primarily had in mind. With $n=2$, the restriction to binary conjunctions makes good sense, and all the examples he discusses are of this kind.

We also mention the results obtained in  \cite{mansfield2012hardy,abramsky2013robust,mansfield2017consequences,simmons}, which study the analogous problem with respect to \emph{possibilistic contextuality}, that is, whether the supports of the probability distributions in the empirical model can be obtained from a set of global assignments. In that case, the complexity of deciding contextuality is shown to be NP-complete in the complexity parameter $k$ in general; a precise delineation is given of the tractability boundary in terms of the values of the parameters. Moreover, as Rui Soares Barbosa has pointed out,\footnote{Personal communication.} if we take $n$ as the complexity parameter, there is a simple algorithm for detecting possibilistic contextuality which is polynomial in the size of the instance. Thus the complexity of detecting possibilistic complexity runs completely in parallel with the probabilistic case.

\section{The ``edge of logical contradiction'' \textit{vs.}~the ``boundary of paradox''}

We give a final quotation from Pitowsky \cite[p.~113]{pitowsky1994george}:
\begin{quotation}
\textit{A violation of Boole's conditions of possible experience cannot be encountered when all the frequencies concerned have been measured on a single
sample. Such a violation simply entails a logical contradiction; `observing' it
would be like `observing' a round square. We expect Boole's conditions to hold
even when the frequencies are measured on distinct large random samples.
But they are systematically violated, and there is no easy way out (see below).
We thus live `on the edge of a logical contradiction'. An interpretation of quantum mechanics,
an attempt to answer the WHY question, is thus an effort to save logic.}
\end{quotation}
In my view, this states the extent of the challenge posed to classical logic by quantum mechanics too strongly. As we have discussed, the observational data predicted by quantum mechanics and confirmed by actual experiments consists of families of probability distributions, each defined on different sample spaces, corresponding to the different contexts. Since the contexts overlap, there are relationships between the sample spaces, which are reflected in coherent relationships between the distributions, in the form of consistent marginals. But there is no ``global'' distribution, defined on a sample space containing all the observable quantities, which accounts for all the empirically observable data. This \emph{does} pose a challenge to the understanding of quantum mechanics as a physical theory, since it implies that we cannot ascribe definite values to the physical quantities being measured, independent of whether or in what context they are measured. It does not, however, challenge classical logic and probability, which can be used to describe exactly this situation.

For this reason, I prefer to speak of contextuality as living ``on the boundary of paradox'' \cite{abramsky2017contextuality}, as a signature of non-classicality, and one for which there is increasing evidence that it plays a fundamental r\^ole in quantum advantage in information-processing tasks \cite{AndersBrowne,RaussendorfSC,Howard2014,abramsky2017contextual}. So this boundary seems likely to prove a fruitful place to be. But we never actually cross the boundary, for exactly the reasons vividly expressed by Pitowsky in the opening two sentences of the above quotation.

There is much more to be said about the connections between logic and contextuality. In particular:
\begin{itemize}
\item There are notions of possibilistic and strong contextuality, and of All-versus-Nothing contextuality, which give a hierarchy of strengths of contextuality, and which can be described in purely logical terms, without reference to probabilities \cite{AbramskyBrandenburger}.
\item These ``possibilistic'' forms of contextuality can be connected with logical paradoxes in the traditional sense. For example, the set of contradictory propositions used in section~\ref{bellsec} to derive Bell's theorem form a ``Liar cycle'' \cite{AbramskyEtAl:ContextualityCohomologyAndParadox}.
\item There is also a topological perspective on these ideas. The four propositions from section~\ref{bellsec} form a discrete version of the M\"obius strip. Sheaf cohomology can be used to detect contextuality \cite{AbramskyEtAl:ContextualityCohomologyAndParadox}.
\item The logical perspective on contextuality leads to the recognition that the same structures arise in many non-quantum areas, including databases \cite{abramsky2013relational}, constraint satisfaction \cite{abramsky2013robust}, and generic inference \cite{abrcaru}.
\item In \cite{abramsky2017contextual}, an inequality of the form $p_F  \geq  \NCF(e)d$ is derived for several different settings involving information-processing tasks. Here $p_F$ is the failure probability, $\NCF(e)$ is the non-contextual fraction of an empirical model viewed as a resource, and $d$ is a parameter measuring the ``difficulty'' of the task. Thus the inequality shows the necessity of increasing the amount of contextuality in the resource in order to increase the success probability. In several cases, including certain games, communication complexity, and shallow circuits, the parameter $d$ is $\frac{n-K}{n}$, where $K$ is the $K$-consistency we encountered in formulating logical Bell inequalities~(\ref{LBell2}).
\end{itemize}

%The limited time available for the preparation of this article has meant that we have not been able to include detailed discussion of these topics. 
We refer to the reader to the papers cited above for additional information on these topics.

\section{Concluding remarks}

Itamar Pitowsky's work on quantum foundations combines lucid analysis, conceptual insights and mathematically sophisticated and elegant results.
We have discussed some recent and ongoing work, and related it to his contributions, which remain a continuing source of inspiration.

\paragraph{Acknowledgements} My thanks to Meir Hemmo and Orly Shenker for giving me the opportunity to contribute to this volume in honour of Itamar Pitowsky. I had the pleasure of meeting Itamar on several occasions when he visited Oxford.
%, and he gave a talk in my seminar on correlation polytopes.

I would also like to thank my collaborators in the work I have described in this paper: Adam Brandenburger, Rui Soares Barbosa, Shane Mansfield, Kohei Kishida, Ray Lal, Giovanni Car\`u, Lucien Hardy, Phokion Kolaitis and Georg Gottlob.
My thanks also to Ehtibar Dzhafarov, whose Contextuality-by-Default  theory \cite{dzhafarov2015contextuality} has much common ground with my own approach, for our ongoing discussions.
\bibliographystyle{plain}
\bibliography{bibfile,refs_cf} 

\begin{thebibliography}{10}

\bibitem{abramsky2013relational}
Samson Abramsky.
\newblock {Relational Databases and Bell's Theorem}.
\newblock In Val Tannen, Limsoon Wong, Leonid Libkin, Wenfei Fan, Wang{-}Chiew
  Tan, and Michael~P. Fourman, editors, {\em In Search of Elegance in the
  Theory and Practice of Computation - Essays Dedicated to Peter Buneman},
  volume 8000 of {\em Lecture Notes in Computer Science}, pages 13--35.
  Springer, 2013.

\bibitem{abramsky2017contextuality}
Samson Abramsky.
\newblock Contextuality: At the borders of paradox.
\newblock In Elaine Landry, editor, {\em Categories for the Working
  Philosopher}. Oxford University Press, 2017.

\bibitem{AbramskyEtAl:ContextualityCohomologyAndParadox}
Samson Abramsky, Rui~Soares Barbosa, Kohei Kishida, Raymond Lal, and Shane
  Mansfield.
\newblock Contextuality, cohomology and paradox.
\newblock In Stephan Kreutzer, editor, {\em 24th EACSL Annual Conference on
  Computer Science Logic (CSL 2015)}, volume~41 of {\em Leibniz International
  Proceedings in Informatics ({LIPIcs})}, pages 211--228, Dagstuhl, Germany,
  2015. Schloss Dagstuhl--Leibniz-Zentrum f{\"u}r Informatik.

\bibitem{abramsky2017contextual}
Samson Abramsky, Rui~Soares Barbosa, and Shane Mansfield.
\newblock Contextual fraction as a measure of contextuality.
\newblock {\em Physical Review Letters}, 119(5):050504, 2017.

\bibitem{AbramskyBrandenburger}
Samson Abramsky and Adam Brandenburger.
\newblock The sheaf-theoretic structure of non-locality and contextuality.
\newblock {\em New J. Phys.}, 13(11):113036, 2011.

\bibitem{abrcaru}
Samson Abramsky and Giovanni Car\`u.
\newblock Non-locality, contextuality and valuation algebras: a general theory
  of disagreement.
\newblock {\em Philosophical Transactions of the Royal Society A}, 2019.
\newblock To appear.

\bibitem{abramsky2013robust}
Samson Abramsky, Georg Gottlob, and Phokion Kolaitis.
\newblock Robust constraint satisfaction and local hidden variables in quantum
  mechanics.
\newblock In {\em Twenty-Third International Joint Conference on Artificial
  Intelligence}, 2013.

\bibitem{AbramskyHardy:LogicalBellIneqs}
Samson Abramsky and Lucien Hardy.
\newblock Logical {B}ell inequalities.
\newblock {\em Phys. Rev. A}, 85(6):062114, 2012.

\bibitem{AndersBrowne}
Janet Anders and Dan~E Browne.
\newblock Computational power of correlations.
\newblock {\em Phys. Rev. Lett.}, 102(5):050502, 2009.

\bibitem{bell1964einstein}
J.S. Bell.
\newblock {On the {E}instein-{P}odolsky-{R}osen paradox}.
\newblock {\em Physics}, 1(3):195--200, 1964.

\bibitem{boole1862theory}
G.~Boole.
\newblock On the theory of probabilities.
\newblock {\em Philosophical Transactions of the Royal Society of London},
  152:225--252, 1862.

\bibitem{christof1997porta}
T.~Christof, A.~L{\"o}bel, and M.~Stoer.
\newblock {PORTA-POlyhedron Representation Transformation Algorithm}.
\newblock {\em Publicly available via ftp://ftp. zib.
  de/pub/Packages/mathprog/polyth/porta}, 1997.

\bibitem{DantzigThapa:LinearProgramming2}
George~B. Dantzig and Mukund~N. Thapa.
\newblock {\em Linear programming 2: {T}heory and extenstions}.
\newblock Springer Series in Operations Research and Financial Engineering.
  Springer Verlag, 2003.

\bibitem{dzhafarov2015contextuality}
Ehtibar~N Dzhafarov, Janne~V Kujala, and Victor~H Cervantes.
\newblock Contextuality-by-default: {A} brief overview of ideas, concepts, and
  terminology.
\newblock In {\em International Symposium on Quantum Interaction}, pages
  12--23. Springer, 2015.

\bibitem{ElitzurPopescuRohrlich1992:QuantumNonlocalityForEachPairInAnEnsemble}
Avshalom~C. Elitzur, Sandu Popescu, and Daniel Rohrlich.
\newblock Quantum nonlocality for each pair in an ensemble.
\newblock {\em Phys. Lett. A}, 162(1):25--28, 1992.

\bibitem{ghirardi1980general}
G.C. Ghirardi, A.~Rimini, and T.~Weber.
\newblock {A general argument against superluminal transmission through the
  quantum mechanical measurement process}.
\newblock {\em Lettere Al Nuovo Cimento (1971--1985)}, 27(10):293--298, 1980.

\bibitem{Howard2014}
Mark Howard, Joel Wallman, Victor Veitch, and Joseph Emerson.
\newblock Contextuality supplies the {`}magic{'} for quantum computation.
\newblock {\em Nature}, 510(7505):351--355, 2014.

\bibitem{jordan1983quantum}
T.F. Jordan.
\newblock {Quantum correlations do not transmit signals}.
\newblock {\em Physics Letters A}, 94(6-7):264, 1983.

\bibitem{mansfield2017consequences}
Shane Mansfield.
\newblock {Consequences and applications of the completeness of Hardy's
  nonlocality}.
\newblock {\em Physical Review A}, 95(2):022122, 2017.

\bibitem{mansfield2012hardy}
Shane Mansfield and Tobias Fritz.
\newblock Hardy's non-locality paradox and possibilistic conditions for
  non-locality.
\newblock {\em Foundations of Physics}, 42(5):709--719, 2012.

\bibitem{pitowsky1989quantum}
Itamar Pitowsky.
\newblock {\em {Quantum Probability, Quantum Logic}}, volume 321 of {\em
  Lecture Notes in Physics}.
\newblock Springer, 1989.

\bibitem{pitowsky1991correlation}
Itamar Pitowsky.
\newblock Correlation polytopes: their geometry and complexity.
\newblock {\em Mathematical Programming}, 50(1-3):395--414, 1991.

\bibitem{pitowsky1994george}
Itamar Pitowsky.
\newblock {George Boole's ``Conditions of Possible Experience'' and the Quantum
  Puzzle}.
\newblock {\em The British Journal for the Philosophy of Science},
  45(1):95--125, 1994.

\bibitem{RaussendorfSC}
Robert Raussendorf.
\newblock Contextuality in measurement-based quantum computation.
\newblock {\em Phys. Rev. A}, 88(2):022322, 2013.

\bibitem{simmons}
Andrew~W Simmons.
\newblock On the computational complexity of detecting possibilistic locality.
\newblock {\em Journal of Logic and Computation}, 28:203--217, 2018.

\bibitem{strichman2002solving}
O.~Strichman.
\newblock {On solving Presburger and linear arithmetic with SAT}.
\newblock In {\em Formal Methods in Computer-Aided Design}, pages 160--170.
  Springer, 2002.

\bibitem{Bonf}
Eric~W. Weisstein.
\newblock Bonferroni inequalities.
\newblock \url{http://mathworld.wolfram.com/BonferroniInequalities.html}.
\newblock MathWorld--A Wolfram Web Resource.

\end{thebibliography}

\end{document}